\documentclass[sigconf,nonacm]{acmart}

\setcopyright{none}
\renewcommand\footnotetextcopyrightpermission[1]{}

\AtBeginDocument{%
  }

\usepackage{mwe}
\usepackage{balance}
\newcommand{\ED}{\operatorname{ED}}

\begin{document}

\title{HiPS: Hierarchical PDF Segmentation of Doctrinal Legal Books}

\author{Sabine Wehnert}
\orcid{0000-0002-5290-0321}
\affiliation{%
   \institution{Ruhr University Bochum, RC Trust}
  \city{Bochum}
  \country{Germany}
}

\author{Harikrishnan Changaramkulath}
\affiliation{%
\institution{Otto von Guericke University}
  \city{Magdeburg}
  \country{Germany}
}

\author{Ivan Habernal}
\orcid{0000-0002-0990-4554}
\affiliation{%
  \institution{Ruhr University Bochum, RC Trust}
  \city{Bochum}
  \country{Germany}
}

\renewcommand{\shortauthors}{Wehnert et al.}

\begin{abstract}
PDF parsers have recently improved on page-level layout understanding. However, recovering a document-global section hierarchy with reliable boundaries remains brittle for deeply structured books: many systems expose only page-local heading roles, assume shallow depth, or rely on high-quality PDF tags or Table of Contents (TOC) metadata, and public gold-standard data for deep book hierarchies is scarce. We present HiPS for hierarchical PDF segmentation of doctrinal legal books and make two main contributions. First, we release a gold-standard benchmark of 49 open-access law books with 9,812 manually curated headings, hierarchy levels, and page anchors, enabling evaluation of title detection, hierarchy reconstruction, and section boundary assignment. Second, we introduce complementary segmentation pipelines: a TOC-based parser for books with reliable outline metadata and a TOC-free LLM-refined pipeline that combines OCR whitespace cues, XML typography, and local context. Across a broad comparison against open-source parsers and multimodal/LLM baselines, the TOC-based pipeline is strongest when metadata is complete, while the LLM-refined pipeline improves heading precision, deep-level recovery, and boundary quality when metadata is missing or noisy.
\end{abstract}



\maketitle

\section{Introduction}
PDFs preserve visual fidelity, but much of their structure remains only weakly machine-readable. Recent benchmarks and surveys show that PDF document parsing systems still emphasize page-level layout analysis or limited structure extraction rather than consistent document-global hierarchy reconstruction across multi-page documents \cite{DBLP:conf/acl/LiA0CZLH00S25,giovannini2025survey,DBLP:journals/corr/abs-2410-21169} (see App.~\ref{sec:related} for extended related work). This limitation has consequences for doctrinal legal books, whose long argumentative exposition is organized through deeply nested section hierarchies. In practice, many systems expose only page-local heading roles, assume shallow hierarchies, or depend on reliable PDF tags or outline metadata~\cite{DBLP:conf/acl/LiA0CZLH00S25,javanmard2025pdf-data-extraction-benchmark}. As a result, downstream legal AI systems can access the text, but not the structure that makes legal exposition navigable and citable.

This gap matters for Legal NLP. Section-aware access to doctrinal books supports retrieval at the right granularity, traceable citation of explanatory passages, and grounded context construction for downstream tasks such as question answering, drafting, and retrieval-augmented generation. Existing structure-extraction benchmarks are informative, but they are not centered on long, publisher-diverse doctrinal legal books with deep hierarchies and imperfect metadata \cite{DBLP:journals/tacl/ArnoldSCGL19,DBLP:conf/bigdataconf/KangPASSBKTE23,DBLP:conf/acl/LiA0CZLH00S25}. HiPS addresses this setting directly by studying three coupled subtasks: \emph{section title detection}, \emph{hierarchy level allocation}, and \emph{section boundary assignment}.

Challenges include that headings in legal books often mix numbered and unnumbered forms, span multiple lines, and vary strongly across publishers. In our corpus, outline metadata frequently under-reports the true hierarchy depth, especially at intermediate and deep levels (App.~\ref{sec:dataset}, Figs.~\ref{fig:sampling}-\ref{fig:comparison}). Good performance therefore requires reconstructing a consistent document-level tree and aligning headings back into the full text stream. 

\section{Task, Dataset, and Baselines}
\begin{figure}[htpb]
  \centering
  \includegraphics[width=\linewidth]{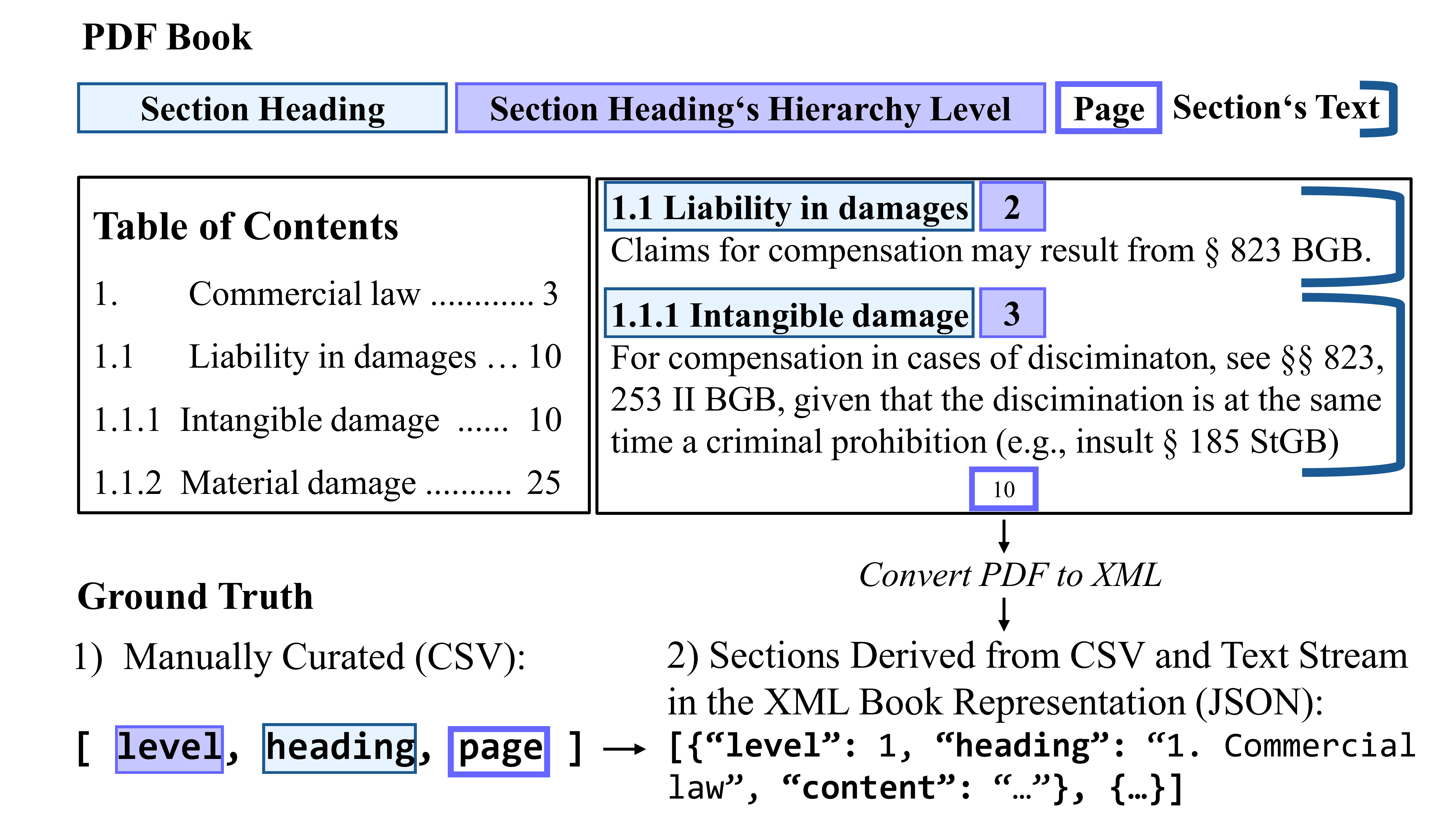}
  \caption{Gold-standard curation procedure and format.}
  \label{fig:gt}
\end{figure}

A central contribution of this work is a gold-standard dataset for deep book hierarchies in the legal domain. We curate 49 open-access law books with 9,812 manually verified headings. For each book, we release heading text, hierarchy level, and page anchor (see Fig.~\ref{fig:gt}).  

From these annotations, we derive section boundaries by aligning the headings to the extracted text stream of the full book. This supports evaluation not only of heading detection, but also of hierarchy reconstruction and end-to-end section segmentation. The dataset fills a gap left by prior resources. While recent datasets and shared tasks cover page-level parsing, TOC extraction, or moderately structured technical documents, they do not capture the combination of deep hierarchy, long-form book structure, incomplete outline metadata, and publisher variability that characterizes doctrinal legal books~\cite{DBLP:conf/icdar/DoucetKCM13,DBLP:conf/bigdataconf/KangPASSBKTE23,DBLP:conf/acl/LiA0CZLH00S25}. In our corpus, several books whose PDF metadata suggests only shallow or mid-depth structure turn out, after manual correction, to contain much deeper hierarchies. This makes the dataset useful not just as a benchmark, but also as evidence that metadata quality is itself part of the problem (App.~\ref{sec:dataset}).

We evaluate systems on three complementary axes. For \emph{title detection}, we measure tolerant precision and recall; for \emph{hierarchy reconstruction}, we compare predicted and gold heading trees with normalized tree edit distance; and for \emph{boundary assignment}, we use established segmentation metrics ($P_k$ and WindowDiff) (App.~\ref{sec:evaluation}). We compare HiPS against a broad set of baselines that reflects the current tool landscape: open-source PDF parsers such as Marker,  Docling, GROBID, and PDFstructure, as well as LLM-based baselines including a zero-shot Llama4-based page parser and several LLM refinement variants with Llama3, Phi4, GPT4 and GPT5 (App.~\ref{sec:baselines}). This breadth matters because current systems fail differently: some over-segment, some flatten the hierarchy, and some recover headings but cannot turn them into faithful section boundaries.

\section{HiPS Overview}
HiPS combines two complementary pipelines. The \textbf{TOC-based PageParser} uses PDF outline metadata when it is reliable and aligns headings to the XML text stream to recover section boundaries. This yields strong performance when the outline is complete and correctly leveled. The \textbf{LLM-refined PageParser} targets books with incomplete or noisy outline metadata. It first extracts heading candidates from OCR whitespace cues and XML features, then asks an instruction-following LLM to clean candidates, decide whether they are true headings, and assign globally consistent hierarchy levels. An alignment stage shared by both pipelines converts predicted headings into full-book section boundaries (App.~\ref{sec:methods} ). The design is grounded in recurring failure modes we observed in the corpus. Heading forms in legal books are highly heterogeneous: some appear without any prefix, others use Arabic numerals, alphabetic markers, Roman numerals, section signs, or spelled-out forms such as "Part Two", and these conventions may be mixed within the same volume. Headings may also span multiple lines and are often signaled more by whitespace and typography than by a stable textual pattern. In our pipelines, OCR whitespace cues recover the visual separation above and below headings, XML features add typography and page-position information, and short local context helps the model suppress running headers, captions, and other false positives. Both pipelines produce the same downstream representation: a hierarchical sequence of headings aligned to contiguous text spans, which makes the output usable for retrieval, citation support, and further knowledge extraction.

\section{Core Results and Contributions}
The evaluation reveals a clear split regime. When outline metadata is complete, the TOC-based PageParser achieves the highest and most stable heading precision and the strongest hierarchy reconstruction. Its median tolerant precision exceeds 0.9, showing that reliable TOC metadata remains best for exact heading recovery. However, recall varies substantially across books because missing TOC entries remove true headings before downstream alignment with the full text even begins (Figs.~\ref{fig:prec_recall} and \ref{fig:etd}).

When metadata is incomplete or noisy, structure-aware LLM refinement is more robust. OCR-augmented prompting reduces false-positive headings compared with XML-only prompting and improves deep-level hierarchy recovery. Among the TOC-free variants, the OCR-enhanced GPT-5 pipeline performs best overall on normalized tree edit distance and on section boundary assignment. The broad baseline comparison also reveals systematic weaknesses of current parsers: Docling attains strong recall but tends to over-segment, GROBID and some Markdown-oriented parsers flatten or truncate the hierarchy, and Llama4 remains unstable on the deepest levels despite its multimodality (Table~\ref{tab:headings-by-level}; Figs.~\ref{fig:segmentation} and \ref{fig:lvl-segmentation}). These findings matter because doctrinal legal books routinely contain 5-7 nested heading levels, precisely where shallow or page-local structure models are lacking.

In sum, the results suggest that hierarchical segmentation of legal books is not solved by either metadata alone or general-purpose multimodal prompting alone. Metadata, layout, and language all carry part of the signal, and HiPS combines them. We share the data, experiments, and further analysis in our repository.\footnote{\url{https://github.com/anybass/HiPS}}

\clearpage
\appendix

\noindent\textit{This appendix supplements the published demo paper with additional references, experimental details, and analysis.}

\section{Extended Related Work}\label{sec:related}
We study structure extraction for \emph{book-length doctrinal legal works} in PDF format, where the goal is to recover a \emph{document-global} hierarchy of headings \emph{and} assign section boundaries across pages.
This differs from page-level layout analysis (e.g., detecting tables/figures and reading order) and text-only topic segmentation, which typically assumes clean sentence streams and does not model PDF layout or book-level hierarchy.

\subsection{Benchmarks and Shared Tasks for Structure Extraction}
A growing set of datasets and shared tasks drives progress in document structure extraction, but their outputs and document domains vary.
WikiSection provides section boundaries for Wikipedia articles and is widely used for evaluating text-based section segmentation \cite{DBLP:journals/tacl/ArnoldSCGL19}.
In finance, the FinTOC shared tasks evaluate title detection and TOC extraction from financial PDFs \cite{DBLP:conf/bigdataconf/KangPASSBKTE23}.
For scanned books, the ICDAR Book Structure Extraction competitions released ground truth for building hyperlinked TOCs from digitized books~\cite{DBLP:conf/icdar/DoucetKCM13}.
READoc formulates structured extraction as an end-to-end task (multi-page PDF to semantically rich Markdown) and provides 3{,}576 PDF-Markdown pairs from arXiv/GitHub/Zenodo together with a unified Standardization-Segmentation-Scoring evaluation suite \cite{DBLP:conf/acl/LiA0CZLH00S25}.
While READoc targets broad technical documents rather than doctrinal legal books, its benchmarks show that systems optimized for page-level layout understanding often fail to produce globally consistent document structure.

These benchmarks are informative, but they generally do not capture the combination of constraints central to our setting: deep (up to level 7) \emph{book hierarchies}, strong publisher variability, incomplete or noisy outline metadata, and end-to-end boundary assignment aligned to in-body headings in long doctrinal legal texts.

\subsection{Outline and Hierarchy Reconstruction}
Beyond assuming a fixed, shallow schema (e.g., the canonical structure of many scientific articles), several approaches aim to reconstruct an explicit outline tree.
Bentabet et al.\ propose a TOC-generation pipeline that detects titles and orders them into a hierarchy without requiring a parsable TOC page, producing a heading tree that mirrors the document outline \cite{DBLP:conf/icdar/BentabetJF19}.
Cao et al.\ introduce HELD \cite{DBLP:journals/jcst/CaoCZL22}, a tree-construction framework for long documents that builds a \emph{variable-depth} logical hierarchy by sequentially inserting each element into a growing tree; in their benchmark, 90\% of headings occur at depths 3--7 and some documents reach depth 11.
They evaluate hierarchy quality using a stricter criterion in which a node is correct only if its entire root-to-node path matches the ground truth, penalizing cases where the depth is correct but the parent chain is wrong.
These approaches are closest in spirit to recovering a document-level tree, but they are commonly evaluated on financial reports or scientific documents and typically emphasize outline recovery rather than boundary assignment over the full text stream in publisher-diverse, book-length doctrinal legal works.

\subsection{From Page-Level Parsing to Document-Global Hierarchies}
Most widely deployed PDF parsers are optimized for \emph{page-level} understanding: they detect layout elements and sometimes label headings, yet often expose page-local roles or shallow heading depths and do not reliably reconstruct a single document-global hierarchy with section boundaries \cite{giovannini2025survey,DBLP:journals/corr/abs-2410-21169}.
Recent work on hierarchical document structure analysis and ``document hierarchy parsing'' introduces datasets and models that connect layout units across pages into structured representations \cite{DBLP:conf/aaai/MaDHZZZL23,DBLP:conf/emnlp/XingCGSYBZY24}.
End-to-end systems such as Detect-Order-Construct jointly perform layout detection, reading-order prediction, and hierarchy construction, reporting strong results on these benchmarks \cite{DBLP:journals/pr/WangHZSH24}.
However, these resources and evaluations largely focus on general technical documents (e.g., papers/reports) with moderate hierarchy depth and more regular heading conventions, and they do not specifically benchmark deeply nested, publisher-variable doctrinal legal books.
This gap motivates our dataset and evaluation: we test whether existing parsers and LLM-based pipelines can recover deep document-global hierarchies and boundaries in doctrinal legal works, and whether structure-aware preprocessing (OCR whitespace cues, XML features, and local context) improves robustness when outline metadata is unreliable.

\section{Dataset Details}\label{sec:dataset}

We are primarily interested in deep book hierarchies, structuring complex topics, which can be often encountered in law books.

\noindent\textbf{Gold standard.}
For each book we provide two ground-truth artifacts (see Fig.~\ref{fig:gt}).
First, a manually curated TOC file as a CSV with rows \texttt{[level, heading, page]}.
Second, a derived segmentation in JSON format whose nodes store \texttt{level}, \texttt{heading}, and the corresponding section \texttt{content}.
Because the JSON contains substantial book text, we do not distribute it for copyright reasons; however, it is fully reproducible from the released CSV annotations and our alignment code that matches the headings in the CSV to the section titles within the fulltext of the XML representation of the book.

\begin{figure}[htpb]
  \centering
\includegraphics[width=\linewidth]{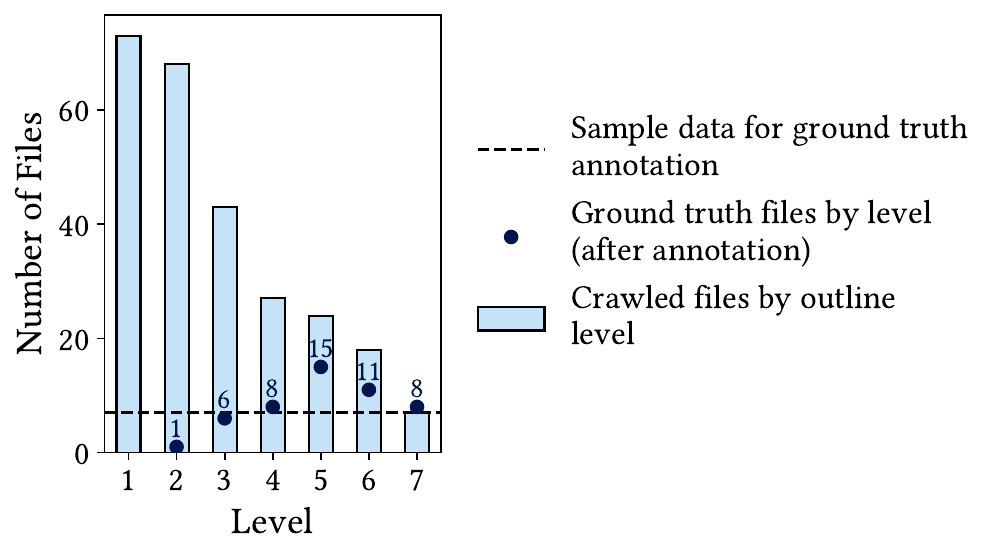}
  \caption{Sampling method for dataset creation. While we sampled an equal number of 7 files per maximum hierarchy depth from 1--7 based on the PDF's outline metadata, their actual distribution changed substantially after ground truth annotation. Many headings from levels above 4 were not included in the outline metadata, which is why now the biggest group with 15 files has a ground truth hierarchy depth of 5. }
\label{fig:sampling}
\end{figure}

As a first step, we crawled all English open access PDF books under the category ``Law'' from the Open Research Library.\footnote{\url{https://openresearchlibrary.org/}} This resulted in a corpus of 362 books, and thereof 259 books had outline metadata for the TOC. These 259 books formed the basis for our data sampling to create ground truth annotations.

Figure~\ref{fig:sampling} illustrates the sampling process based on the data distribution. With this data, we choose to work in a rule-based way, using the PDF that has been converted to XML with the \texttt{pdftohtml} tool inside the Poppler library.\footnote{\url{poppler.freedesktop.org}} That library also extracts the TOC metadata and stores them along with the page in the outline information. In that way, information such as font size and positioning can be used for the matching of headings inside the fulltext on the indicated page for finding the boundaries.

Annotations were produced by a single annotator following written guidelines; all headings were validated by ensuring they could be matched in the extracted text stream. We report this as a limitation and release the data to enable corrections and extensions.

The following annotation guideline was employed for creating the ground truth data:
\begin{itemize}
    \item We took the outline metadata (=TOC) of the PDFs as a basis for the ground truth data. In case the headings were properly included in the TOC, we could directly use this information, including the page numbers and hierarchy level.
    \item We examined the PDF files closely for any missing text suggesting to be a section heading. This involved the book title, which was sometimes included in the metadata, but not in the majority of cases. If a section heading was missing or misspelled in the TOC metadata (so that it would not match the section title within the referenced page in the fulltext), it was adjusted and included into the ground truth data along with its page and hierarchy level.
    \item We used the core logic from the \texttt{TOC-Based PageParser} (see Section \ref{sec:boundaries}) to detect all ground truth TOC data inside the book's fulltext. With that we could ensure that all ground truth headings were detectable in the fulltext, i.e., there were no unmatched headings logged. This step was necessary because we compare against the XML files: There may be OCR errors or headings split across different XML text nodes, requiring advanced matching strategies. 
\end{itemize}

\begin{figure}[htpb]
    \centering
    \includegraphics[width=\linewidth]{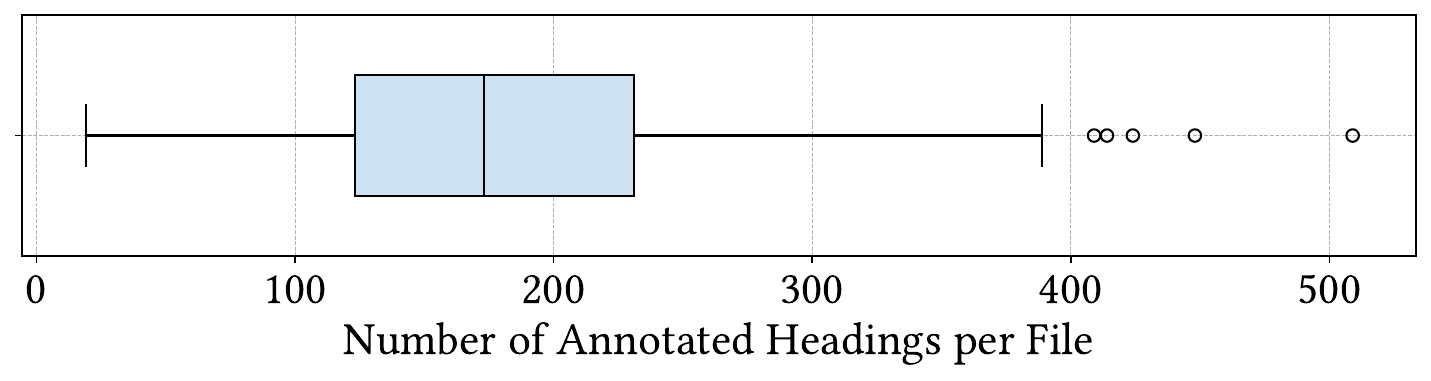}
    \caption{Distribution of annotated headings per book (ground-truth TOC).}
    \label{fig:annotation}
\end{figure}

The ground truth data was produced through manual verification and correction of the mistakes found in the original TOC metadata. This involved checking for missing, misaligned, or incorrectly leveled headings, cross-referencing against the fulltext, and adjusting as needed to reflect the true structural segmentation. Figure \ref{fig:annotation} illustrates the distribution of annotated ground truth headings in each book file. In total, we manually curated 9,812 headings inside the ground truth corpus across the 49 files.

\begin{figure}[ht]
  \centering
\includegraphics[width=\linewidth]{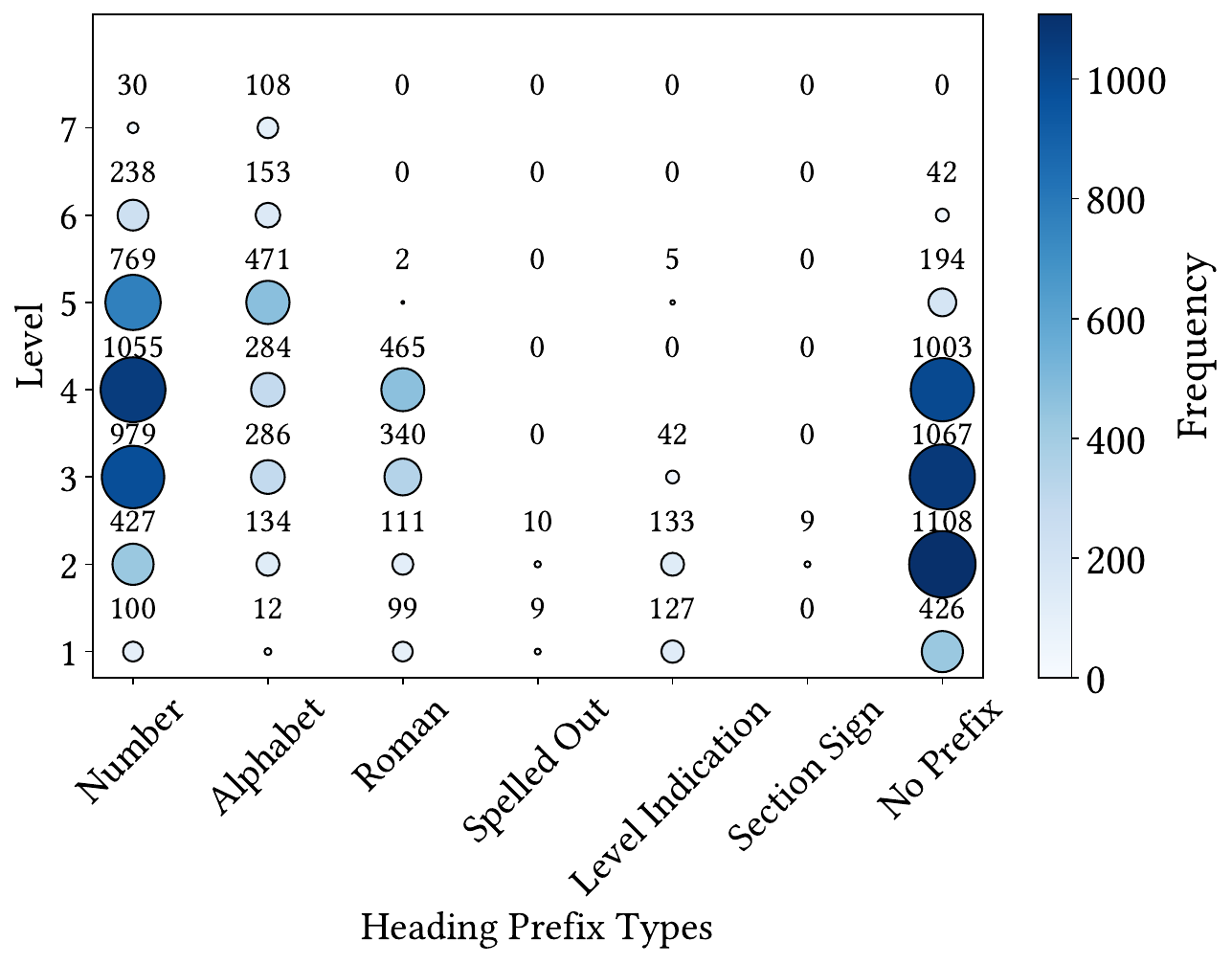}
  \caption{Heading prefix distribution within the ground truth dataset. It depicts a multi-label distribution of all heading prefixes from the ground truth data, with the majority being no prefix at all (just text). This makes it difficult to use only rule-based approaches to detect headings, since there may be no reliable pattern. However, the second most frequent category is some form of number prefix (``1.'', ``1)'', ``1''), which potentially has a finite number of permutations. In only 9 headings we found a section sign (``\S'') and in 19 cases we had a spelled out prefix number (e.g., ``Part Two''), which was often combined with a level indication (``Chapter'', ``Part'', ``Appendix'', ...). Alphabetic prefixes were frequently encountered throughout almost all levels, roman numerals only in levels lower than 6.}
\label{fig:prefix}
\end{figure}
\begin{figure}[ht]
  \centering
\includegraphics[width=\linewidth]{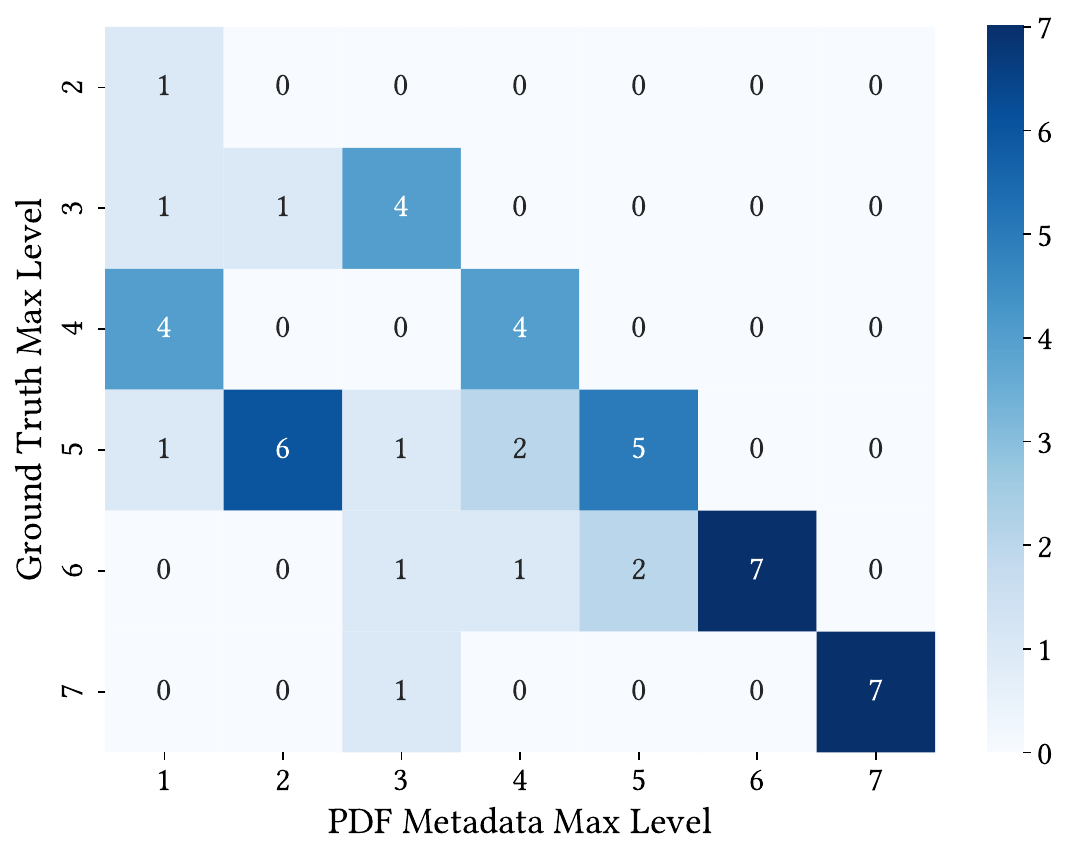}
  \caption{Comparison between maximum hierarchy depth based on PDF outline metadata and the ground truth. For example, a cluster of 6 files marked with maximum depth 2 in the metadata were found to have actual depths of 5 in the ground truth. Similarly, metadata levels 3 and 4 were associated with a wide spread of actual depths, indicating that TOC metadata alone is often insufficiently reliable for accurately assessing structural depth. On the other hand, metadata indicating maximum depths of 6 and 7 were aligned with their annotated counterparts, suggesting greater consistency at the deeper end of the spectrum. }
\label{fig:comparison}
\end{figure}

Figure \ref{fig:prefix} illustrates the diversity of formatting options authors have chosen for headings. The two most common types are number prefixes or no prefixes at all. The latter makes the task of heading detection harder for text-only methods that do not rely on structural or visual cues.
Figure~\ref{fig:comparison} visualizes the relationship between the maximum hierarchy levels found in the PDF outline metadata and those verified through manual annotation. The distribution reveals several mismatches between the actual structure and the metadata of the PDF. Given the higher quality of TOCs in deeply structured books, we hypothesize that the TOC-Based PageParser achieves strong performance at these levels.

\section{Detailed Method Description}\label{sec:methods}
We formulate three subtasks in order to perform a hierarchical section-wise book segmentation:
\begin{itemize}
    \item \textbf{Section Title Detection: } First, text spans in the PDF corresponding to section headings are identified. For this, often Table of Contents metadata is helpful, but we noticed in Appendix~\ref{sec:dataset} that for deeper hierarchy levels, the section titles are often omitted by the authors/editors. Therefore, different techniques to detect headings are applied in this work.
    \item \textbf{Hierarchy Level Allocation: } Second, hierarchy levels are assigned for all section titles. In order to analyze how well LLMs can distinguish between different hierarchy levels among sections, we investigate whether they can reproduce the original hierarchy. In this sense, the level as a numeric value is not as important as the assignment of consistent values to distinguish parent and child section titles.
    \item \textbf{Section Boundary Assignment: } Third, headings are aligned to the extracted text stream, and successive headings delimit section content. This yields a full segmentation of the book into hierarchical sections.
\end{itemize}

We evaluate each component against the gold standard (App.~\ref{sec:dataset}).
In practice, performance depends strongly on the availability and quality of Table of Contents (TOC) metadata.
We therefore propose two complementary approaches:
\begin{itemize}
    \item \textbf{TOC-Based PageParser} uses TOC metadata when it is reliable, and
    \item \textbf{LLM-Refined PageParser} recovers structure even when TOC metadata is incomplete or noisy.
\end{itemize}

\subsection{Section Title Detection and Hierarchy Level Allocation}
In this section, we describe several approaches employed for the task of detecting section titles and identifying their respective hierarchy level. We implement two approaches (TOC-Based PageParser and LLM-Refined PageParser) and compare them to representative open-source parsers and LLM baselines (App.~\ref{sec:baselines}).

\subsubsection{TOC-Based PageParser}\label{sec:tocpp}
This approach extracts section titles, logical pages, and hierarchy levels directly from the TOC metadata of the book PDF files. TOC entries often differ slightly from full-text headings due to typographical inconsistencies, formatting variations, or Optical Character Recognition (OCR) artifacts. Our method accounts for these discrepancies using multiple matching strategies during heading-to-text alignment (detailed in App.~\ref{sec:boundaries}). When the TOC metadata is complete and accurate, this approach yields a faithful deep hierarchy without requiring learned or resource-intensive LLM-based heading detection.

\subsubsection{LLM-Refined PageParser}\label{sec:llmpp}
When TOC metadata is missing, incomplete, or unreliable, we use an LLM-refined pipeline.
The method first extracts heading \emph{candidates} using layout cues and then refines them with a Large Language Model (LLM) during title detection and level assignment.

\begin{table*}[htpb]
\centering
\caption{Comparison of representative tools/APIs for section-level parsing. ``Global hierarchy'' refers to constructing a single document-level tree (beyond page-local labels).}
\label{tab:tools-comparison}
\small
\begin{tabular}{p{2.2cm} p{3.4cm} p{3.1cm} p{2.5cm} p{1.7cm} p{1cm} p{1cm}}
\toprule
\textbf{Tool / API} & \textbf{Max heading levels exposed} & \textbf{Global hierarchy reconstruction} & \textbf{Section boundary assignment} & \textbf{TOC reconciliation} & \textbf{Open-source} & \textbf{Base-line} \\
\midrule
\texttt{Google Cloud Document AI} & up to \texttt{heading-5} & No (page-local labels) & No & No & No & No \\
\texttt{Azure AI Document Intelligence} & \textit{roles} (e.g., sectionHeading) & No (page-local labels) & No & No & No & No \\
\texttt{Adobe PDF Extract API} & depends on PDF tags & Yes, if tagged; otherwise limited & No & No & No & No \\
\texttt{Marker} & implicit via Markdown \# levels & Partial (derivable) & No (text-only unless post-processed) & No & Yes & Yes \\
\texttt{GROBID} & shallow-to-moderate (TEI \texttt{div/head}) & Yes (nested TEI), but domain-tuned to papers & No (requires post-processing) & No & Yes & Yes \\
\texttt{Docling} & unlimited (JSON/MD headers) & Partial (derivable) & No (requires post-processing) & No & Yes & Yes \\
\texttt{PDFstructure} & n/a (heuristic) & Partial (heuristic grouping) & Yes (heuristic spans) & No & Yes & Yes \\
\bottomrule
\end{tabular}
\end{table*}

\paragraph{Candidate Selection}
We tested two candidate-selection strategies: XML-derived features and OCR with contextual features.

\noindent\textbf{XML Features:}
Text elements are extracted from the XML representation of the book, along with metadata such as font size, bold/italic formatting, page number, and positional attributes. This metadata will then be supplied to the LLM for selected title candidates. The selection process is based on font usage statistics, prioritizing fonts that appear less frequently in the document but still within a reasonable threshold. While some publishers use distinct properties such as larger font sizes or discrete font styles for headings, there are scenarios where they utilize homogeneous font sizes and styles. Moreover, XML features can fail to capture contextual layout information, such as multi-line headings, leading to high false-positive rates. A key insight from our analysis is that headings are often distinguished by whitespace surrounding the heading text (empty lines above/below), which makes them visually separable from body text; leveraging this property from XML alone is challenging.

\noindent\textbf{XML-OCR Features:}
 Given the limitations of relying solely on XML-based features, we adopted Tesseract OCR\footnote{\url{https://github.com/madmaze/pytesseract}} for processing PDF pages as images and detecting headings based on spatial context. Each page is rendered at 300 DPI and OCR returns text with layout information. Our approach specifically identifies headings by detecting lines that are either preceded or followed by empty lines, serving as an indicator of spatial separation. This method effectively captures both single-line and multi-line headings. We further refine these detections through additional filtering, partially based on XML features, to exclude false positives. This includes removing lines with excessive word counts or those exhibiting specific punctuation patterns (e.g., headings terminating with commas or colons).

\paragraph{Title Detection and Level Assignment.}
For each candidate heading, we provide the LLM with layout features (e.g., page number, font ID, text height/width, page position) and a short trailing text passage. We instruct the models to remove noise,
such as fragmented text or special characters, and predict whether candidates are true headings based on the features. Additionally, the prompt instructs to assign realistic hierarchy levels (typically 2--3 levels, up to 10) by analyzing semantic relevance and contextual relationships, maintaining consistency with previously identified headings. We evaluate several instruction-following LLMs: \texttt{Llama3}, \texttt{Phi4}, \texttt{GPT-4}, and \texttt{GPT-5}.

\subsection{Baselines and Tool Coverage}\label{sec:baselines}
We compare our approaches to widely used open-source parsers that emit headings from PDFs.
Table~\ref{tab:tools-comparison} summarizes representative tools/APIs and motivates the baseline selection: most systems either provide page-local labels, assume shallow hierarchies, or omit end-to-end boundary assignment. We do not include proprietary cloud services as baselines, including Google Cloud Document AI,\footnote{\url{https://cloud.google.com/document-ai}} Azure AI Document Intelligence,\footnote{\url{https://learn.microsoft.com/azure/ai-services/document-intelligence/}} and Adobe PDF Extract API,\footnote{\url{https://developer.adobe.com/document-services/apis/pdf-extract/}} because they are closed-source commercial systems and thus do not support transparent, fully reproducible evaluation in our setting.

\paragraph{Marker, Docling and GROBID}
Marker,\footnote{\url{https://github.com/datalab-to/marker}} Docling,\footnote{\url{https://github.com/IBM/docling}} and GROBID\footnote{\url{https://github.com/kermitt2/grobid}} output headings in Markdown/JSON/TEI formats (sometimes with implicit nesting), enabling derivation of a hierarchy.
They do not assign section boundaries end-to-end; therefore, when evaluating boundaries, we apply our shared heading-to-text alignment procedure (App.~\ref{sec:boundaries}).

\subsubsection{PDFstructure}
Among the structural PDF parsing tools, we emphasize that \texttt{PDFstructure}\footnote{\url{https://github.com/ChrizH/pdfstructure}} is the only \textit{end-to-end} parser baseline: It is built on top of the Python library pdfminer.six\footnote{\url{https://github.com/pdfminer/pdfminer.six}} and uses several heuristics for parsing the PDFs without making use of the TOC, yielding a representation of section titles, their level, and the corresponding section's text. \texttt{PDFstructure} uses the following features for its heuristics: text size, line margins and spacing, font properties, as well as spatial relationships among text boxes.

\subsubsection{Llama4 (Zero-shot)}
Our multimodal baseline uses a \texttt{vLLM} deployment of \texttt{Llama4} as a prompt-only, vision-based page parser.\footnote{\url{https://huggingface.co/meta-llama/Llama-4-Maverick-17B-128E-Instruct-FP8}} Each PDF page is rendered as a 250 DPI image and processed independently. For each page, the model is prompted to classify the page type (narrative content, navigational list, or publication metadata) and to extract only structural headings (e.g., Part/Chapter/Section), while ignoring running headers/footers, captions, marginalia, references, and TOC-style entries. The model outputs headings as CSV triples \texttt{(hierarchy\_level, heading\_text, page\_number)}. We then run a second cleanup pass over the aggregated outputs, prompting the model to remove duplicates, merge split headings within a page, and enforce globally consistent hierarchy levels, including a consistent mapping from dotted numeric prefix depth to hierarchy level.

\subsection{Hierarchical Section Boundary Assignment}\label{sec:boundaries}
We segment each book into TOC-aligned sections using a rule-based, single-pass
procedure that preserves TOC hierarchy levels. TOC entries are ordered by page,
filtered to remove empty or placeholder headings, and indexed so that each page
has an ordered list of expected headings. We then scan the XML document page by page,
maintaining a forward cursor within each page to avoid duplicating text when
multiple headings occur on the same page.

If no heading is expected on a page, we append the page text (in reading order) to
the current section. Otherwise, for each expected heading we search forward from
the cursor to place the next boundary. We first attempt a fine-grained match on
individual text fragments; if no match is found, we fall back to matching on
reconstructed reading-order lines to capture headings that span multiple
fragments, and map the matched line back to its fragment span. When a boundary is
found, text before it is committed to the current section, a new section is
opened with the TOC-provided heading and level, and scanning continues after the
match. To avoid dropping adjacent content, we retain the matched fragment/span
unless it appears to be a standalone heading line. Unmatched headings are logged
for diagnostics but do not interrupt processing.

Heading matching is designed to be robust to OCR and layout variation. We
normalize both the candidate text and the heading by deleting extra whitespace,
removing spacing around separators (e.g., ``/'' and ``-''), collapsing spurious
inter-character spacing (e.g., ``P U B L I C'' to ``PUBLIC''), stripping
punctuation, and lowercasing. A heading is accepted if any of the following holds: exact equality after removing spaces, substring containment after
removing spaces, or fuzzy similarity using the \texttt{fuzzywuzzy} partial
ratio\footnote{\url{https://github.com/seatgeek/thefuzz}} with a threshold of 80. If
the fuzzy score is below the threshold, we apply a conservative word-overlap fallback
that requires at least one heading word of length $\ge 4$ to appear in the candidate text. After scanning all pages, we emit the trailing open section and merge immediately consecutive sections with the same heading and
level, remove empty stubs, and report unmatched TOC headings. This procedure is used for all pipelines, except for \texttt{PDFstructure} which already provides sectioned output.

\section{Extended Experimental Setup and Results}\label{sec:evaluation}
We perform most of our experiments on an AMD Ryzen 9 3900X (24) \@ 3.800G CPU equipped with a NVIDIA GeForce RTX 3090 GPU.\footnote{Except for the Llama 4 experiments on 6$\times$ NVIDIA H200 GPUs (140 GB each)} The implementation is available online\footnote{\url{https://github.com/anybass/HiPS}} and consists of Python scripts and Jupyter notebooks.

\subsection{Section Title Detection}
We measure the performance of detecting the corresponding section titles with edit-distance-tolerant Precision $P_{ED}$ and Recall $R_{ED}$.
The modifications towards more tolerance were necessary, since there may be small inaccuracies due to trailing control characters or whitespace that the respective methods did not accommodate. For each document $d$, let $G_d$ denote the set of ground-truth section titles and
$\hat{G}_d$ the set of predicted section titles. Let $\ED(x,y)$ be the character-level Levenshtein edit distance\footnote{\url{https://github.com/rapidfuzz/Levenshtein}} between strings $x$ and $y$,
and let $\mathbb{I}[\cdot]$ be the indicator function, i.e., $\mathbb{I}[\text{true}]=1$ and $\mathbb{I}[\text{false}]=0$.
We count a predicted title $\hat{g}\in\hat{G}_d$ as a true positive if there exists a ground-truth title $g\in G_d$ such that
$\ED(\hat{g},g)\le 2$ (analogously for recall):

\begin{align*}
P_{ED}(d)=\frac{1}{|\hat{G}_d|}\sum_{\hat{g}\in \hat{G}_d}
\mathbb{I}\!\left[\min_{g\in G_d}\ED(\hat{g},g)\le 2\right]
, \text{ and} \\
R_{ED}(d)=\frac{1}{|G_d|}\sum_{g\in G_d}
\mathbb{I}\!\left[\min_{\hat{g}\in \hat{G}_d}\ED(g,\hat{g})\le 2\right].
\end{align*}

\begin{figure*}[htpb]
  \centering
  \includegraphics[width=\textwidth]{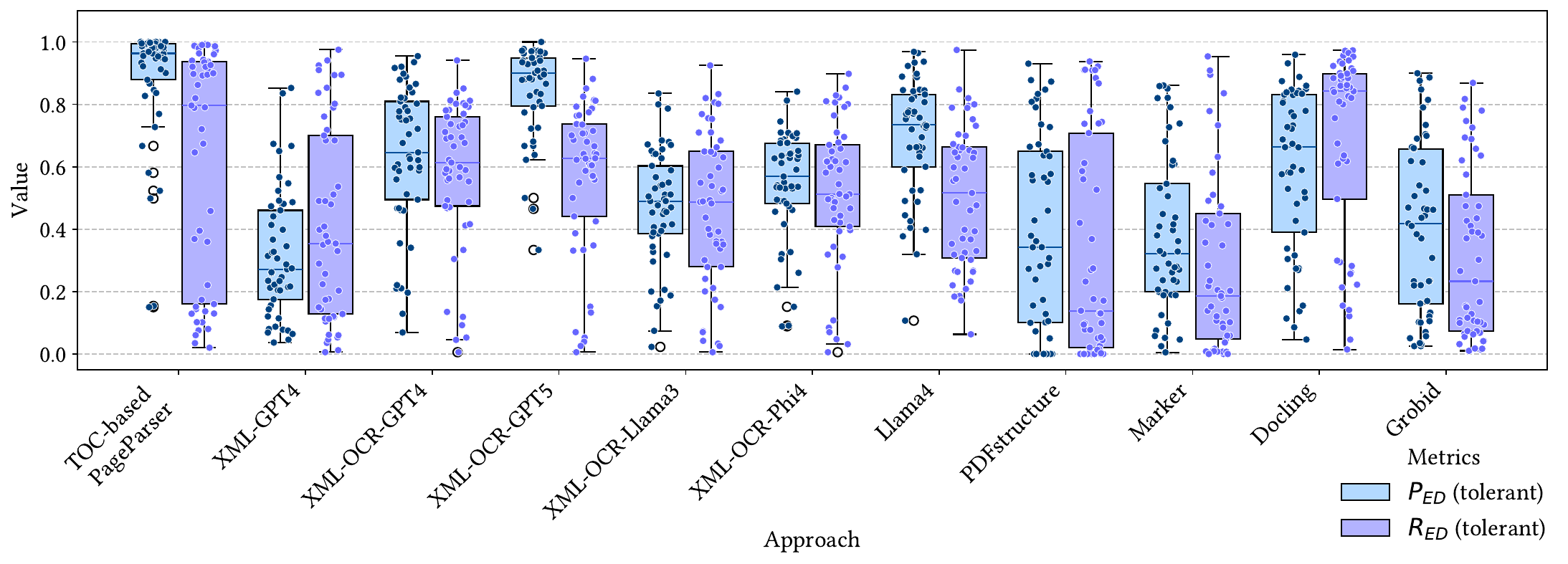}
  \caption{Section title detection results using edit-distance-tolerant precision $P_{ED}$ and recall $R_{ED}$ (higher is better; a predicted title matches gold if $\mathrm{ED}\le2$ after normalization). Boxplots show per-book distributions across 49 doctrinal legal books (median line, IQR box, whiskers). \texttt{TOC-based PageParser} achieves the highest and most stable precision ($P_{ED} \gtrsim 0.9$ median), but its recall varies strongly across books, reflecting dependence on TOC metadata completeness (missing deep-level entries reduce $R_{ED}$). Among TOC-free methods, OCR-augmented LLM pipelines reduce false positives compared to XML-only prompting: \texttt{XML-GPT4} over-detects headings (low $P_{ED}$), while \texttt{XML-OCR-GPT5} yields the best precision among non-TOC approaches. \texttt{Docling} attains the strongest recall ($R_{ED} \gtrsim 0.8$ median) but at lower precision, consistent with over-segmentation.}
  \label{fig:prec_recall}
\end{figure*}

\begin{figure*}[ht]
  \centering
  \includegraphics[width=\textwidth]{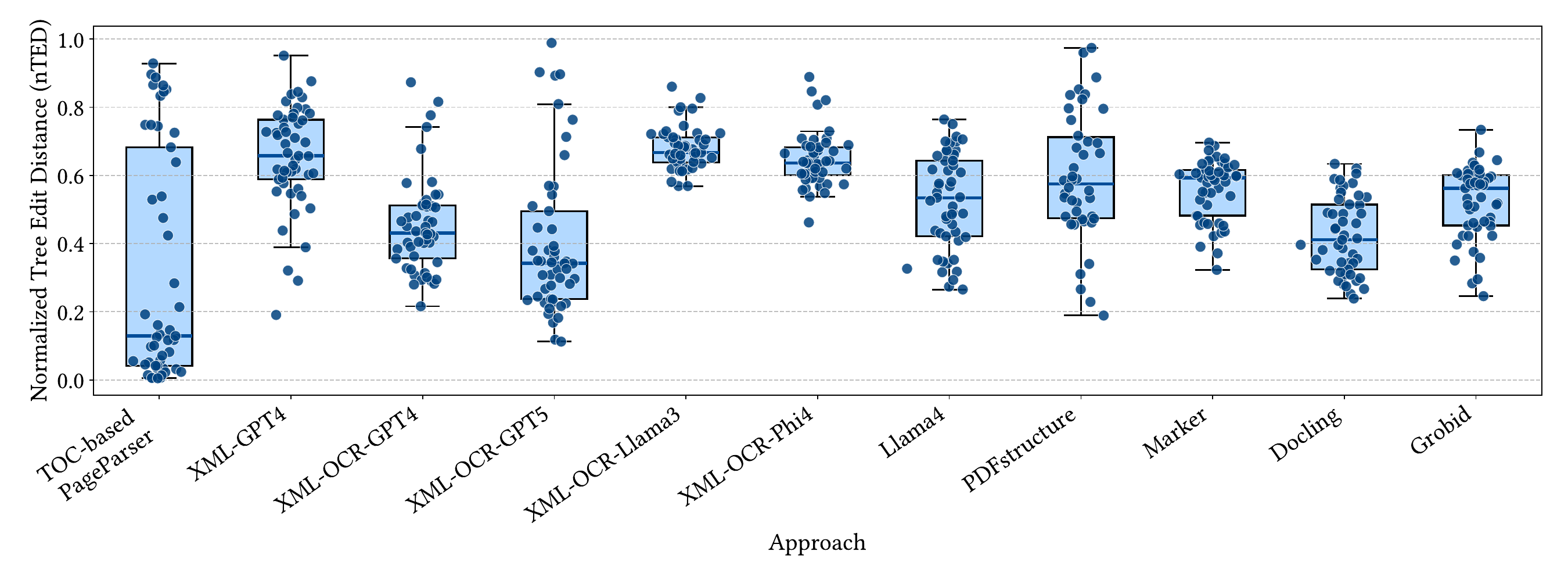}
  \caption{Hierarchy level allocation accuracy measured by normalized Tree Edit Distance (nTED) between the predicted and gold heading trees (lower is better). Boxplots show per-book distributions across 49 doctrinal legal books. \texttt{TOC-based PageParser} achieves the lowest median nTED (around 0.1), but with high variance when TOC metadata is incomplete, reflecting that missing or misleveled TOC entries distort the induced tree. Among TOC-free methods, \texttt{XML-OCR-GPT5} yields the best median performance (around 0.3), indicating improved recovery of parent--child relations at deeper levels. In contrast, \texttt{XML-GPT4} (no OCR) and open-source LLM variants (\texttt{XML-OCR-Llama3}, \texttt{XML-OCR-Phi4}) show the highest edit rates (median nTED $>0.6$). \texttt{Docling} collapses headings into a single level (no levels beyond L1) and \texttt{Grobid} produces only a small number of level-2 headings; their nTED is therefore dominated by title matching and shallow structure rather than faithful multi-level reconstruction.}
  \label{fig:etd}
\end{figure*}

Figure~\ref{fig:prec_recall} summarizes title-detection accuracy across all books using tolerant precision and recall. Overall, \texttt{TOC-based PageParser} yields the highest precision, but recall depends on TOC completeness, while OCR-augmented LLM pipelines improve precision relative to XML-only prompting, and \texttt{Docling} achieves the strongest recall. Note that the errors of the title detection will propagate to subsequent tasks.

\subsection{Section Title Hierarchy Level Allocation}
Evaluating hierarchy level allocation is non-trivial because level identifiers (e.g., 1,2,3) are not meaningful as numeric values; instead, correctness is determined by whether the induced parent-child and sibling relations among section titles match the reference hierarchy. We therefore model each document's table of contents as an \emph{ordered, rooted tree} $T=(V,E)$, where each node $v\in V$ corresponds to a section title and edges represent the parent-child relation implied by the allocated levels (a synthetic root is added for convenience). We compare a predicted tree $T_{\text{pred}}$ to the ground-truth tree $T_{\text{gt}}$ using the Zhang-Shasha \emph{tree edit distance} (TED),\footnote{\url{https://github.com/timtadh/zhang-shasha}} defined as the minimum-cost sequence of edit operations that transforms one tree into the other. The allowed elementary operations are node insertion, node deletion, and node relabeling (substitution), under the constraint that ancestor relations and sibling order are preserved  \cite{DBLP:journals/siamcomp/ZhangS89}. TED is
\begin{equation}
\mathrm{TED}(T_{\text{pred}},T_{\text{gt}})=\min_{S\in\mathcal{S}(T_{\text{pred}},T_{\text{gt}})} \sum_{e\in S} c(e),
\end{equation}
where $\mathcal{S}$ is the set of valid edit scripts and $c(e)$ is the cost of operation $e$.
To emphasize structural relationships over numeric level values, we use unit costs for structural edits. Insertions and deletions each cost 1. We compare node labels using normalized section titles: we strip leading/trailing whitespace, replace newlines with spaces and collapse repeated whitespace, remove a small set of punctuation characters commonly found in PDF-extracted TOCs (e.g., typographic quotes and parentheses), and escape backslashes to ensure robust parsing. After this normalization, aligning two nodes incurs cost 0 if the normalized titles are identical and cost 1 otherwise. Because raw TED grows with the number of nodes (longer TOCs provide more opportunities for edits), we report a size-normalized distance instead:
\begin{equation}
\mathrm{nTED}(T_{\text{pred}},T_{\text{gt}})=\frac{\mathrm{TED}(T_{\text{pred}},T_{\text{gt}})}{|V_{\text{pred}}|+|V_{\text{gt}}|},
\end{equation}
where $|V|$ counts section-title nodes (excluding the synthetic root). With unit insert/delete costs, $|V_{\text{pred}}|+|V_{\text{gt}}|$ is an upper bound on the edits required (delete all nodes then insert all nodes), making nTED interpretable as an ``edit rate'' per node and comparable across documents of different TOC lengths. Traditional normalizations such as $\mathrm{TED}/(|V_1|+|V_2|)$ and $\mathrm{TED}/\max(|V_1|,|V_2|)$ are commonly discussed in literature \cite{DBLP:journals/fcsc/LiZ11,DBLP:conf/emnlp/HwangLYKS21}. Intuitively, nTED is the fraction of section nodes that must be inserted, deleted, or reattached (via edits) to make the predicted hierarchy match the ground truth.

Figure~\ref{fig:etd} reports hierarchy reconstruction accuracy using nTED. Overall, \texttt{TOC-based PageParser} performs best when TOC metadata is complete, while \texttt{XML-OCR-GPT5} is the strongest TOC-free approach; systems that collapse the hierarchy into one or two levels (e.g., \texttt{Docling}, \texttt{Grobid}) cannot faithfully recover deep structure.

Although numeric level labels are not meaningful in isolation, comparing heading counts per reference level offers a complementary diagnostic of depth allocation: it exposes systematic flattening or premature truncation of the hierarchy, which in turn manifests as higher nTED. When comparing the mean number of predicted section titles per file at each level against the ground truth (GT), we observe that level~7 is accurately recovered only by \texttt{TOC-based PageParser} and \texttt{XML-OCR-GPT5}. Most other approaches substantially underestimate the number of headings in documents that contain at least one level~7 heading in the GT (Table~\ref{tab:headings-by-level}). Beyond level~7, several methods exhibit systematic structural biases: \texttt{XML-GPT4} markedly overproduces level~2 headings, whereas \texttt{Marker} concentrates predictions in levels~3--4, while \texttt{Llama4} is overpredicting shallow headings (level~2).

\begin{table}[htpb!]
\small
\centering
\caption{Mean number of headings per document by predicted hierarchy level (L1 is the shallowest). This diagnostic complements nTED by showing whether a method \emph{flattens} the hierarchy (mass in L1--L2), \emph{truncates} depth (zeros at deeper levels), or recovers deep levels. Our two approaches are grouped together: the TOC-based PageParser preserves deep levels when TOC metadata is reliable but undercounts intermediate levels when TOC entries are missing; the LLM-refined PageParser without OCR (XML--GPT4) over-produces shallow headings (especially L2), while XML-OCR-based candidate selection reduces shallow false positives and enables deeper recovery. Among baselines, Docling and GROBID largely collapse headings into L1 (no depth), Marker recovers mid-level headings but truncates beyond L4, and Llama4 yields mostly shallow levels with no L6--L7 output.}
\label{tab:headings-by-level}
\begin{tabular}{lrrrrrrr}
\toprule
Model & L1 & L2 & L3 & L4 & L5 & L6 & L7 \\
\midrule
GT & 12.51 & 33.71 & 59.40 & 66.90 & 42.44 & 22.79 & 17.25 \\
\midrule
\multicolumn{8}{l}{\textit{Proposed approaches}} \\
\textbf{TOC-b. P.} & 12.76 & 22.96 & 38.88 & 40.62 & 27.24 & 19.21 & 17.12 \\
\addlinespace
\multicolumn{8}{l}{\textbf{LLM-r.\ P.\ variants}} \\
XML-GPT4 & 98.18 & 149.82 & 65.02 & 12.24 & 2.68 & 1.00 & 0.00 \\
X.-O.-Ll.3 & 51.76 & 58.18 & 40.17 & 25.14 & 14.53 & 3.42 & 2.12 \\
X.-O.-Phi4 & 49.02 & 60.35 & 43.02 & 21.90 & 10.68 & 2.05 & 0.88 \\
X.-O.-GPT4 & 22.82 & 64.78 & 63.08 & 16.40 & 5.68 & 3.63 & 7.12 \\
X.-O.-GPT5 & 5.96  & 22.27 & 39.00 & 29.52 & 13.03 & 12.16 & 18.62 \\
\midrule
\multicolumn{8}{l}{\textit{Baselines}} \\
Docling    & 241.82 & 0.00 & 0.00 & 0.00 & 0.00 & 0.00 & 0.00 \\
Grobid     & 170.06 & 13.12 & 0.00 & 0.00 & 0.00 & 0.00 & 0.00 \\
Marker     & 56.94 & 41.78 & 62.10 & 73.36 & 0.00 & 0.00 & 0.00 \\
PDFstruc.  & 12.95 & 30.67 & 62.39 & 43.11 & 27.67 & 20.47 & 11.57 \\
Llama4     & 50.37 & 50.53 & 20.19 & 4.93 & 0.50 & 0.00 & 0.00 \\
\bottomrule
\end{tabular}
\end{table}

\subsection{Hierarchical Section Boundary Assignment}\label{subsec:wdpk}
We evaluate predicted section boundaries against the ground truth using two standard text segmentation
metrics, $P_k$~\cite{DBLP:journals/ml/BeefermanBL99} and \emph{WindowDiff}~\cite{DBLP:journals/coling/PevznerH02}. Both slide a window of length $k$ over the sequence of sentence units and quantify boundary disagreements
(lower is better). Let a document be a sequence of $N$ sentence units. Let $S^{\star}$ denote the gold segmentation and $\hat{S}$ the predicted segmentation. For any segmentation $S$, define the boundary indicator
\begin{equation}
\delta_S(i,j)=
\begin{cases}
1,& \exists\,\text{a boundary between } i \text{ and } j,\\
0,& \text{otherwise,}
\end{cases}
\label{eq:delta}
\end{equation}
for indices $1\le i<j\le N$.
Given a window size $k$, $P_k$~\cite{DBLP:journals/ml/BeefermanBL99} is computed by sliding the window and
comparing whether the two endpoints are separated by a boundary:
\begin{equation}
P_k(S^{\star},\hat{S})
=\frac{1}{N-k}\sum_{i=1}^{N-k}
\mathbb{I}\!\left[\delta_{S^{\star}}(i,i+k)\neq \delta_{\hat{S}}(i,i+k)\right].
\end{equation}

For \emph{WindowDiff}, let $b_S(i,j)$ be the number of segment boundaries between indices $i$ and $j$ under segmentation $S$. Then
\begin{equation}
\mathrm{WD}(S^{\star},\hat{S})
=\frac{1}{N-k}\sum_{i=1}^{N-k}
\mathbb{I}\!\left[b_{S^{\star}}(i,i+k)\neq b_{\hat{S}}(i,i+k)\right],
\end{equation}
i.e., an error is counted if the predicted and gold segmentations contain a different
number of boundaries in a window.

\begin{figure*}[htbp!]
  \centering
  \includegraphics[width=\textwidth]{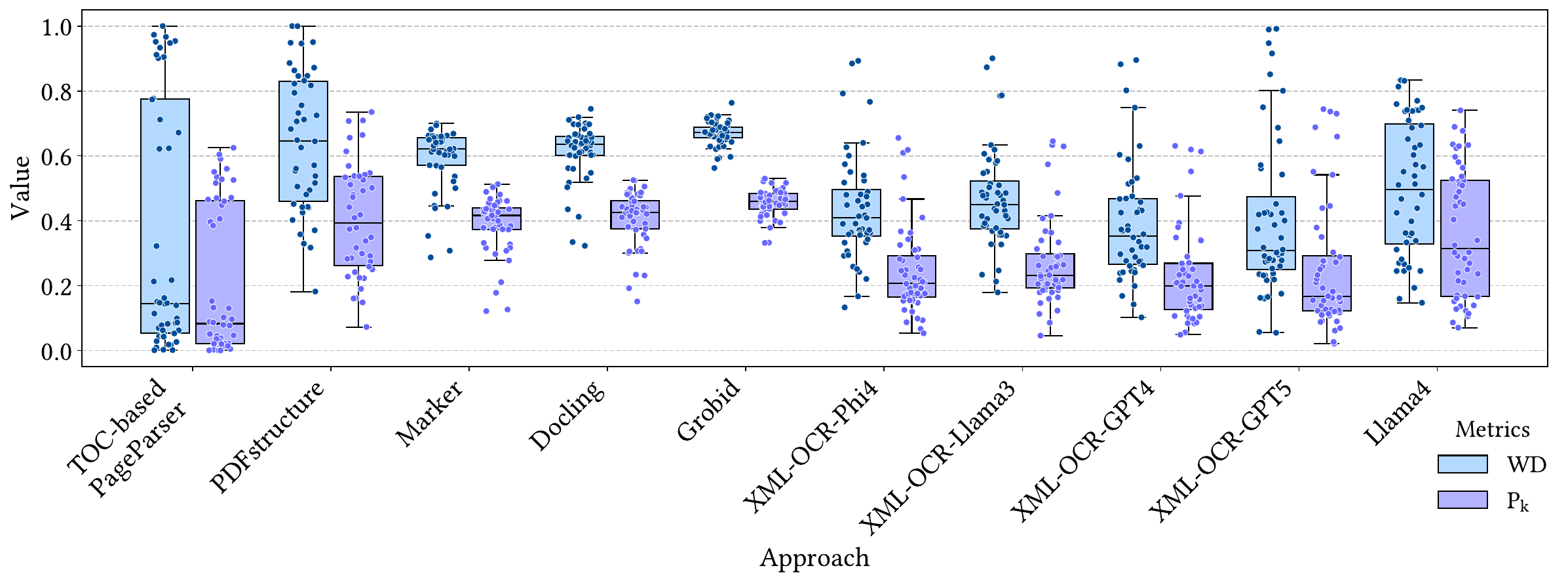}
  \caption{Hierarchical section boundary assignment evaluated with \emph{WindowDiff} (WD) and $P_k$ (lower is better), aggregated across 49 doctrinal legal books. \texttt{TOC-based PageParser} achieves the lowest median error on both metrics but shows high variance across books, reflecting dependence on the completeness and accuracy of Table of Contents (TOC) metadata (missing deep-level TOC entries lead to under-segmentation and boundary shifts; cf.\ Fig.~\ref{fig:comparison}). Among TOC-free methods, \texttt{XML-OCR-GPT5} yields the strongest and most stable performance, followed by \texttt{XML-OCR-GPT4}, while several open-source parsers exhibit higher error due to over- or under-segmentation.}
  \label{fig:segmentation}
\end{figure*}

\begin{figure*}[htbp!]
  \centering
  \includegraphics[width=\textwidth]{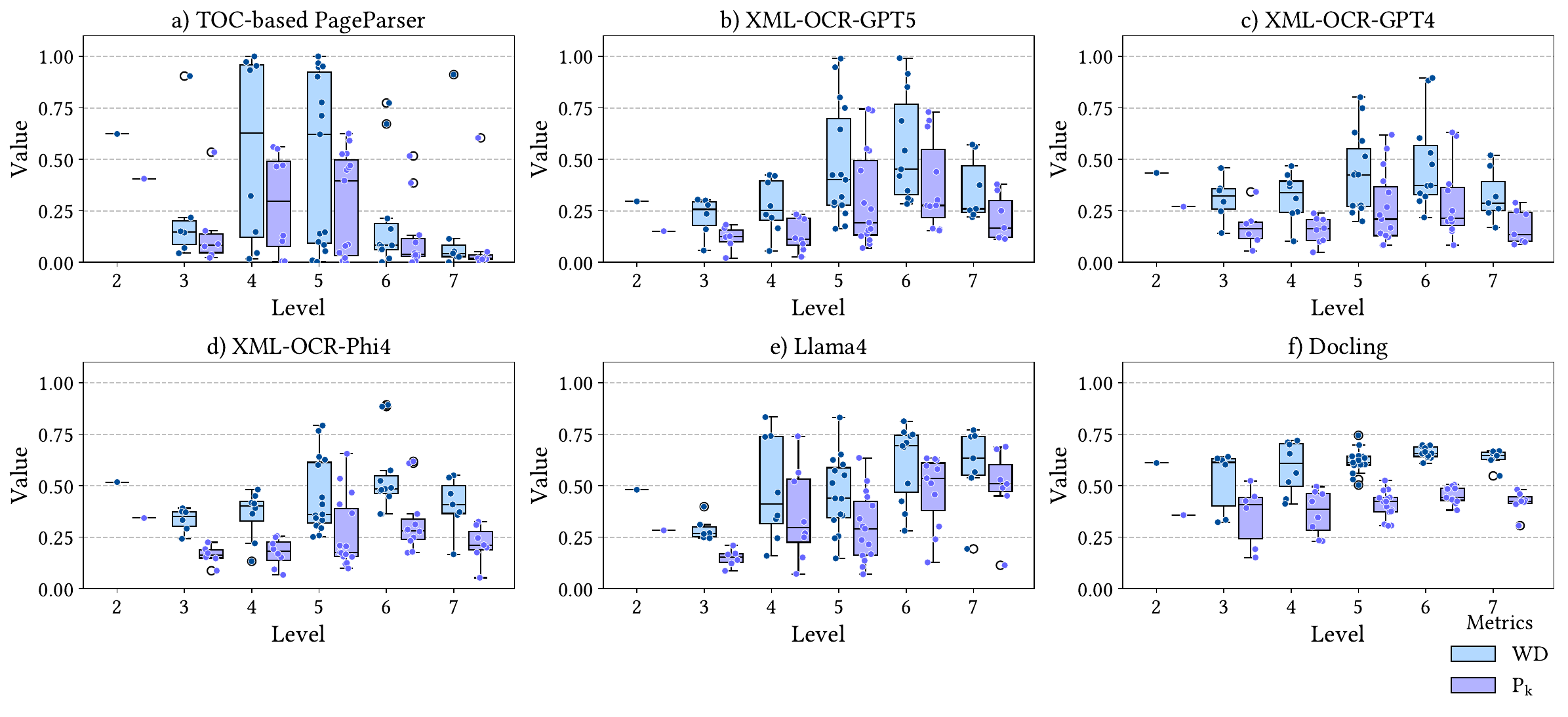}
 \caption{Boundary assignment error by hierarchy level for selected models, reported with WindowDiff (WD) and $P_k$ (lower is better). Panels (a--f) compare complementary regimes: (a) \texttt{TOC-based PageParser} degrades at mid-depth levels (L4--L5) when Table of Contents (TOC) metadata is incomplete, consistent with missing TOC entries that remove expected boundaries; (b,c) \texttt{XML-OCR-GPT5} and \texttt{XML-OCR-GPT4} recover deep levels without using the TOC and maintain comparatively low error across levels, though they tend to under-segment (fewer boundaries than gold), especially at deeper levels; (d) \texttt{XML-OCR-Phi4} yields competitive boundary scores despite weaker hierarchy-level allocation; (e) the vision-only \texttt{Llama4} baseline struggles on the deepest levels (L6--L7); and (f) \texttt{Docling} exhibits elevated error consistent with over-segmentation (too many boundaries).}
  \label{fig:lvl-segmentation}
\end{figure*}

Figure~\ref{fig:segmentation} summarizes boundary assignment performance across models; we next analyze error patterns by hierarchy level and metadata quality.
Figure~\ref{fig:lvl-segmentation} breaks boundary errors down by hierarchy level for representative models and highlights regime-specific failure patterns (metadata dependence vs.\ TOC-free recovery).
Overall, TOC-guided parsing is strongest when TOC metadata is complete, whereas OCR-augmented LLM pipelines are more robust when TOC metadata is missing or noisy.
This suggests that there is no free lunch---even for state-of-the-art LLMs---and future efforts are necessary.

\section{Limitations}
Several factors may influence the validity of our experimental results. In our dataset, we only included PDFs with a single-column layout, and we also did not consider historical books where OCR issues may influence the process substantially. Although our sampling was random and based on the PDFs' TOC metadata, we observed that missing annotated levels are common in books. Thus, our dataset is naturally imbalanced, with fewer files having a maximum hierarchy level below 3. The ground truth was created by a single annotator, and while all ground truth headings were matched in the full text, some hierarchy assignments may still be open to interpretation. We welcome issue reports via our repository,\footnote{\url{https://github.com/anybass/HiPS}} where we plan to release future corrections and extensions.

\section{Ethical Considerations}
We published a dataset of section titles, page numbers, and hierarchy levels from English law books in the Open Research Library. The dataset is intended for non-commercial research and educational use. We designed it to minimize inclusion of expressive content: it consists primarily of factual/functional bibliographic and navigational information, and the extracted headings are necessary for boundary identification. Where copyright or related rights could be implicated, we rely, where applicable, on relevant exceptions and limitations (including, where relevant, text-and-data-mining/research and teaching exceptions). We acknowledge the original authors and publishers (see acknowledgment and source references in our repository). If any issues arise, we are open to addressing them.

\balance
\bibliographystyle{ACM-Reference-Format}
\bibliography{bibliography}

@String{Computer = "{IEEE} Computer" }

@inproceedings{DBLP:conf/emnlp/HwangLYKS21,
  author       = {Wonseok Hwang and
                  Hyunji Lee and
                  Jinyeong Yim and
                  Geewook Kim and
                  Minjoon Seo},
  editor       = {Marie{-}Francine Moens and
                  Xuanjing Huang and
                  Lucia Specia and
                  Scott Wen{-}tau Yih},
  title        = {Cost-effective End-to-end Information Extraction for Semi-structured
                  Document Images},
  booktitle    = {Proceedings of the 2021 Conference on Empirical Methods in Natural
                  Language Processing, {EMNLP} 2021, Virtual Event / Punta Cana, Dominican
                  Republic, 7-11 November, 2021},
  pages        = {3375--3383},
  publisher    = {Association for Computational Linguistics},
  year         = {2021},
  url          = {https://doi.org/10.18653/v1/2021.emnlp-main.271},
  doi          = {10.18653/V1/2021.EMNLP-MAIN.271},
  bibsource    = {dblp computer science bibliography, https://dblp.org}
}

@article{DBLP:journals/fcsc/LiZ11,
  author       = {Yujian Li and
                  Chenguang Zhang},
  title        = {A metric normalization of tree edit distance},
  journal      = {Frontiers Comput. Sci. China},
  volume       = {5},
  number       = {1},
  pages        = {119--125},
  year         = {2011},
  url          = {https://doi.org/10.1007/s11704-011-9336-2},
  doi          = {10.1007/S11704-011-9336-2},
  bibsource    = {dblp computer science bibliography, https://dblp.org}
}

@article{DBLP:journals/ml/BeefermanBL99,
  author    = {Doug Beeferman and
               Adam L. Berger and
               John D. Lafferty},
  title     = {Statistical Models for Text Segmentation},
  journal   = {Mach. Learn.},
  volume    = {34},
  number    = {1-3},
  pages     = {177--210},
  year      = {1999},
  bibsource = {dblp computer science bibliography, https://dblp.org}
}

@article{DBLP:journals/coling/PevznerH02,
  author    = {Lev Pevzner and
               Marti A. Hearst},
  title     = {A Critique and Improvement of an Evaluation Metric for Text Segmentation},
  journal   = {Comput. Linguistics},
  volume    = {28},
  number    = {1},
  pages     = {19--36},
  year      = {2002},
  bibsource = {dblp computer science bibliography, https://dblp.org}
}

@inproceedings{DBLP:conf/bigdataconf/KangPASSBKTE23,
  author       = {Juyeon Kang and
                  Mauli Mehulkumar Patel and
                  Anushka Agrawal and
                  Simhadri Sevitha and
                  Srinivasa Ravi and
                  Sandra Bellato and
                  M. Anand Kumar and
                  Ngawang Dempa Tsang and
                  Mo El{-}Haj},
  editor       = {Jingrui He and
                  Themis Palpanas and
                  Xiaohua Hu and
                  Alfredo Cuzzocrea and
                  Dejing Dou and
                  Dominik Slezak and
                  Wei Wang and
                  Aleksandra Gruca and
                  Jerry Chun{-}Wei Lin and
                  Rakesh Agrawal},
  title        = {Advancements in Financial Document Structure Extraction: Insights
                  from Five Years of FinTOC {(2019-2023)}},
  booktitle    = {{IEEE} International Conference on Big Data, BigData 2023, Sorrento,
                  Italy, December 15-18, 2023},
  pages        = {2839--2844},
  publisher    = {{IEEE}},
  year         = {2023},
  url          = {https://doi.org/10.1109/BigData59044.2023.10386125},
  doi          = {10.1109/BIGDATA59044.2023.10386125},
  bibsource    = {dblp computer science bibliography, https://dblp.org}
}

@article{DBLP:journals/tacl/ArnoldSCGL19,
  author       = {Sebastian Arnold and
                  Rudolf Schneider and
                  Philippe Cudr{\'{e}}{-}Mauroux and
                  Felix A. Gers and
                  Alexander L{\"{o}}ser},
  title        = {{SECTOR:} {A} Neural Model for Coherent Topic Segmentation and Classification},
  journal      = {Trans. Assoc. Comput. Linguistics},
  volume       = {7},
  pages        = {169--184},
  year         = {2019},
  url          = {https://doi.org/10.1162/tacl\_a\_00261},
  doi          = {10.1162/TACL\_A\_00261},
  bibsource    = {dblp computer science bibliography, https://dblp.org}
}

@article{DBLP:journals/pr/WangHZSH24,
  author       = {Jiawei Wang and
                  Kai Hu and
                  Zhuoyao Zhong and
                  Lei Sun and
                  Qiang Huo},
  title        = {Detect-order-construct: {A} tree construction based approach for hierarchical
                  document structure analysis},
  journal      = {Pattern Recognit.},
  volume       = {156},
  pages        = {110836},
  year         = {2024},
  url          = {https://doi.org/10.1016/j.patcog.2024.110836},
  doi          = {10.1016/J.PATCOG.2024.110836},
  bibsource    = {dblp computer science bibliography, https://dblp.org}
}

@inproceedings{giovannini2025survey,
  title={A Survey on Reading Order, Table of Contents, and Structure Extraction in Document Analysis},
  author={Giovannini, Simone and Marinai, Simone},
  booktitle={Proceedings of the IEEE/CVF International Conference on Computer Vision},
  pages={7585--7594},
  year={2025}
}

@article{DBLP:journals/jcst/CaoCZL22,
  author       = {Rongyu Cao and
                  Yixuan Cao and
                  Ganbin Zhou and
                  Ping Luo},
  title        = {Extracting Variable-Depth Logical Document Hierarchy from Long Documents:
                  Method, Evaluation, and Application},
  journal      = {J. Comput. Sci. Technol.},
  volume       = {37},
  number       = {3},
  pages        = {699--718},
  year         = {2022},
  url          = {https://doi.org/10.1007/s11390-021-1076-7},
  doi          = {10.1007/S11390-021-1076-7},
  bibsource    = {dblp computer science bibliography, https://dblp.org}
}

@inproceedings{DBLP:conf/icdar/BentabetJF19,
  author    = {Najah{-}Imane Bentabet and
              others},
  title     = {Table-of-Contents Generation on Contemporary Documents},
  booktitle = {{ICDAR} 2019, Sydney, Australia, September 20-25, 2019},
  pages     = {100--107},
  publisher = {{IEEE}},
  year      = {2019},
  bibsource = {dblp computer science bibliography, https://dblp.org}
}

@inproceedings{DBLP:conf/emnlp/XingCGSYBZY24,
  author       = {Hangdi Xing and
                  Changxu Cheng and
                  Feiyu Gao and
                  Zirui Shao and
                  Zhi Yu and
                  Jiajun Bu and
                  Qi Zheng and
                  Cong Yao},
  editor       = {Yaser Al{-}Onaizan and
                  Mohit Bansal and
                  Yun{-}Nung Chen},
  title        = {DocHieNet: {A} Large and Diverse Dataset for Document Hierarchy Parsing},
  booktitle    = {Proceedings of the 2024 Conference on Empirical Methods in Natural
                  Language Processing, {EMNLP} 2024, Miami, FL, USA, November 12-16,
                  2024},
  pages        = {1129--1142},
  publisher    = {Association for Computational Linguistics},
  year         = {2024},
  url          = {https://doi.org/10.18653/v1/2024.emnlp-main.65},
  doi          = {10.18653/V1/2024.EMNLP-MAIN.65},
  bibsource    = {dblp computer science bibliography, https://dblp.org}
}

@inproceedings{DBLP:conf/aaai/MaDHZZZL23,
  author       = {Jiefeng Ma and
                  Jun Du and
                  Pengfei Hu and
                  Zhenrong Zhang and
                  Jianshu Zhang and
                  Huihui Zhu and
                  Cong Liu},
  editor       = {Brian Williams and
                  Yiling Chen and
                  Jennifer Neville},
  title        = {HRDoc: Dataset and Baseline Method toward Hierarchical Reconstruction
                  of Document Structures},
  booktitle    = {Thirty-Seventh {AAAI} Conference on Artificial Intelligence, {AAAI}
                  2023, Thirty-Fifth Conference on Innovative Applications of Artificial
                  Intelligence, {IAAI} 2023, Thirteenth Symposium on Educational Advances
                  in Artificial Intelligence, {EAAI} 2023, Washington, DC, USA, February
                  7-14, 2023},
  pages        = {1870--1877},
  publisher    = {{AAAI} Press},
  year         = {2023},
  url          = {https://doi.org/10.1609/aaai.v37i2.25277},
  doi          = {10.1609/AAAI.V37I2.25277},
  bibsource    = {dblp computer science bibliography, https://dblp.org}
}

@inproceedings{DBLP:conf/acl/LiA0CZLH00S25,
  author       = {Zichao Li and
                  Aizier Abulaiti and
                  Yaojie Lu and
                  Xuanang Chen and
                  Jia Zheng and
                  Hongyu Lin and
                  Xianpei Han and
                  Shanshan Jiang and
                  Bin Dong and
                  Le Sun},
  editor       = {Wanxiang Che and
                  Joyce Nabende and
                  Ekaterina Shutova and
                  Mohammad Taher Pilehvar},
  title        = {READoc: {A} Unified Benchmark for Realistic Document Structured Extraction},
  booktitle    = {Findings of the Association for Computational Linguistics, {ACL} 2025,
                  Vienna, Austria, July 27 - August 1, 2025},
  series       = {Findings of {ACL}},
  volume       = {{ACL} 2025},
  pages        = {21889--21905},
  publisher    = {Association for Computational Linguistics},
  year         = {2025},
  url          = {https://aclanthology.org/2025.findings-acl.1128/},
  bibsource    = {dblp computer science bibliography, https://dblp.org}
}

@inproceedings{DBLP:conf/icdar/DoucetKCM13,
  author       = {Antoine Doucet and
                  Gabriella Kazai and
                  Sebastian Colutto and
                  G{\"{u}}nter M{\"{u}}hlberger},
  title        = {{ICDAR} 2013 Competition on Book Structure Extraction},
  booktitle    = {12th International Conference on Document Analysis and Recognition,
                  {ICDAR} 2013, Washington, DC, USA, August 25-28, 2013},
  pages        = {1438--1443},
  publisher    = {{IEEE} Computer Society},
  year         = {2013},
  url          = {https://doi.org/10.1109/ICDAR.2013.290},
  doi          = {10.1109/ICDAR.2013.290},
  bibsource    = {dblp computer science bibliography, https://dblp.org}
}

@article{DBLP:journals/siamcomp/ZhangS89,
  author       = {Kaizhong Zhang and
                  Dennis E. Shasha},
  title        = {Simple Fast Algorithms for the Editing Distance Between Trees and
                  Related Problems},
  journal      = {{SIAM} J. Comput.},
  volume       = {18},
  number       = {6},
  pages        = {1245--1262},
  year         = {1989},
  bibsource    = {dblp computer science bibliography, https://dblp.org}
}

@misc{javanmard2025pdf-data-extraction-benchmark,
  author       = {Arash Javanmard},
  title        = {PDF Data Extraction Benchmark 2025: Comparing Docling, Unstructured, and LlamaParse for Document Processing Pipelines},
  howpublished = {Procycons Blog},
  year         = {2025},
  month        = mar,
  note         = {Published 2025-03-24. Accessed 2026-01-31},
  url          = {https://procycons.com/en/blogs/pdf-data-extraction-benchmark/}
}

@article{DBLP:journals/corr/abs-2410-21169,
  author       = {Qintong Zhang and
                  Victor Shea{-}Jay Huang and
                  Bin Wang and
                  Junyuan Zhang and
                  Zhengren Wang and
                  Hao Liang and
                  Shawn Wang and
                  Matthieu Lin and
                  Conghui He and
                  Wentao Zhang},
  title        = {Document Parsing Unveiled: Techniques, Challenges, and Prospects for
                  Structured Information Extraction},
  journal      = {CoRR},
  volume       = {abs/2410.21169},
  year         = {2024},
  url          = {https://doi.org/10.48550/arXiv.2410.21169},
  doi          = {10.48550/ARXIV.2410.21169},
  eprinttype    = {arXiv},
  eprint       = {2410.21169},
  bibsource    = {dblp computer science bibliography, https://dblp.org}
}

\end{document}